\documentclass[aps,prb,reprint,superscriptaddress]{revtex4-1}
\usepackage{siunitx}
\usepackage{graphicx}			
\usepackage{color}

\begin{document}
\newcommand{\NHK}{[(NH$_4$)$_{1-x}$K$_x$]$_2$[FeCl$_5$(H$_2$O)]}
\newcommand{\fivepercent}{[(NH$_4$)$_{1-x}$K$_x$]$_2$[FeCl$_5$(H$_2$O)] with $x = $ 0.06}
\newcommand{\fifteenpercent}{[(NH$_4$)$_{1-x}$K$_x$]$_2$[FeCl$_5$(H$_2$O)] with $x = $ 0.15}
\newcommand{\sixtypercent}{[(NH$_4$)$_{1-x}$K$_x$]$_2$[FeCl$_5$(H$_2$O)] with $x = $ 0.45}
\newcommand{\pureNH}{(NH$_4$)$_2$[FeCl$_5$(H$_2$O)]}
\newcommand{\ND}{(ND$_4$)$_2$[FeCl$_5$(D$_2$O)]}
\newcommand{\pureK}{K$_2$[FeCl$_5$(H$_2$O)]}
\newcommand{\avec}{\textbf{\textit{a}}}
\newcommand{\bvec}{\textbf{\textit{b}}}
\newcommand{\cvec}{\textbf{\textit{c}}}
\newcommand{\Hvec}{\textbf{\textit{H}}}
\newcommand{\Pola}{\textit{P}$_a$}
\newcommand{\Polb}{\textit{P}$_b$}
\newcommand{\Polc}{\textit{P}$_c$}

\title{Magnetoelectric coupling in the mixed erythrosiderite \NHK}


\author{Daniel Br\"uning}
\affiliation{II. Physikalisches Institut, Universit\"at zu K\"oln, Z\"ulpicher Stra{\ss}e 77, 50937 K\"oln, Germany}
\author{Tobias Fr\"ohlich}
\affiliation{II. Physikalisches Institut, Universit\"at zu K\"oln, Z\"ulpicher Stra{\ss}e 77, 50937 K\"oln, Germany}
\author{Malte Langenbach}
\affiliation{II. Physikalisches Institut, Universit\"at zu K\"oln, Z\"ulpicher Stra{\ss}e 77, 50937 K\"oln, Germany}
\author{Thomas Leich}
\affiliation{II. Physikalisches Institut, Universit\"at zu K\"oln, Z\"ulpicher Stra{\ss}e 77, 50937 K\"oln, Germany}
\author{Martin Meven}
\affiliation{RWTH Aachen University, Institut f\"ur Kristallographie, 52056 Aachen, Germany}
\affiliation{J\"ulich Centre for Neutron Science JCNS at Heinz Maier-Leibnitz Zentrum (MLZ), Forschungszentrum J\"ulich GmbH, Lichtenbergstraße 1, 85747 Garching, Germany}
\author{Petra Becker}
\affiliation{Abteilung Kristallographie, Institut f\"ur Geologie und Mineralogie, Universit\"at zu K\"oln, Z\"ulpicher Stra{\ss}e 49b, 50674 K\"oln, Germany}
\author{Ladislav Bohat\'y}
\affiliation{Abteilung Kristallographie, Institut f\"ur Geologie und Mineralogie, Universit\"at zu K\"oln, Z\"ulpicher Stra{\ss}e 49b, 50674 K\"oln, Germany}
\author{Markus Gr\"uninger}
\affiliation{II. Physikalisches Institut, Universit\"at zu K\"oln, Z\"ulpicher Stra{\ss}e 77, 50937 K\"oln, Germany}
\author{Markus Braden}
\affiliation{II. Physikalisches Institut, Universit\"at zu K\"oln, Z\"ulpicher Stra{\ss}e 77, 50937 K\"oln, Germany}
\author{Thomas Lorenz}
\affiliation{II. Physikalisches Institut, Universit\"at zu K\"oln, Z\"ulpicher Stra{\ss}e 77, 50937 K\"oln, Germany}


\date{\today}

\begin{abstract}
We present a study of the dielectric, structural, and magnetic properties of the multiferroic or linear magnetoelectric substitution series \NHK. Pyroelectric currents, magnetic susceptibilities, and thermodynamic properties were examined on large single crystals of the erythrosiderite compounds and detailed magnetic-field versus temperature phase diagrams are derived for three different substitution levels. With increasing potassium concentration the material is tuned from a multiferroic ($x \le 0.06$) to a linear magnetoelectric ($x \ge 0.15$) ground state. In contrast to the respective pure parent compounds with $x=0$ or 1, however, the ferroelectric or linear magnetoelectric polarization in none of the substituted samples is switchable by external electric fields, because these samples exhibit a significant electric polarization already above the magnetic ordering transition. The polarization arises at a higher-lying structural phase transition that is examined by THz spectroscopy and, on a deuterated pure single crystal, by comprehensive neutron-diffraction experiments. The structural phase transition is attributed to an ordering of NH$_4^+$ tetrahedra but does not break inversion symmetry in the pure material, while a finite K content causes pyroelectricity. 

\end{abstract}


\maketitle

\section{Introduction}
During the last decade spin-driven multiferroics with complex antiferromagnetic order, e.g., spiral or cycloidal spin structures, inspired intensive research activities \cite{Spaldin2019, Dong2015, Wang2009, Nan2008, Cheong2007, Fiebig2005}. These magnetic ordering phenomena can induce a ferroelectric polarization, thus ferroelectric and magnetic order are strongly coupled. These so-called type-II multiferroics are improper ferroelectrics; their polarization is rather weak but can be strongly influenced by external magnetic fields. Studies on multiple spin-driven multiferroic materials, e.g., TbMnO$_3$ \cite{Kimura2003}, Ni$_3$V$_2$O$_8$ \cite{Lawes2005}, MnWO$_4$ \cite{Heyer2006,Arkenbout2006}, and NaFe$X_2$O$_6$ ($X=$ Si, Ge) \cite{Kim2012,Jodlauk2007}, determined the underlying coupling mechanisms. Most members show two magnetic ordering transitions with a paraelectric intermediate phase with incommensurate and collinear order and a low-temperature ferroelectric phase arising from an inversion-symmetry breaking magnetic structure.

\begin{figure}
	\includegraphics[width=0.48\textwidth]{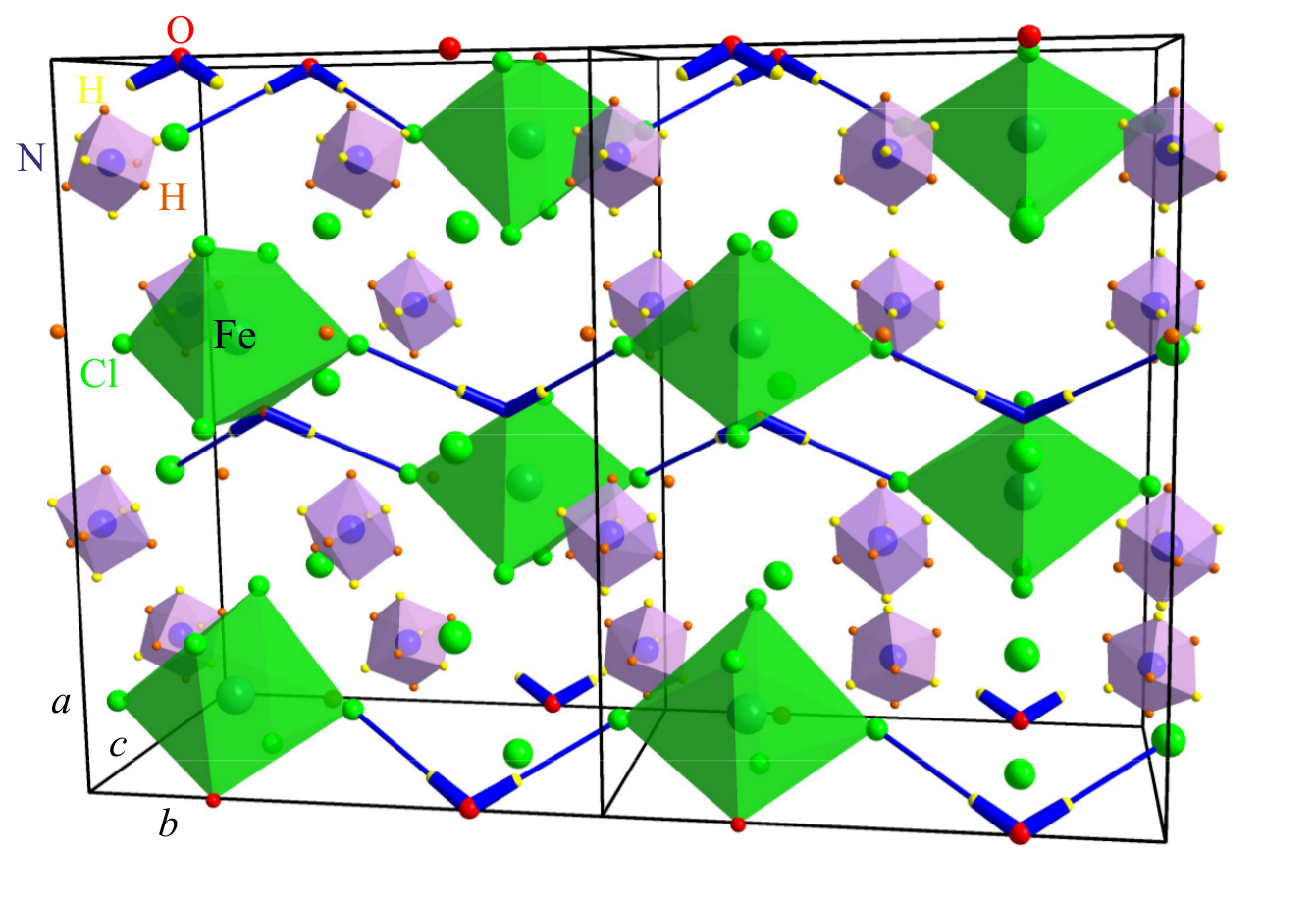}
	\caption{High-temperature structure of \pureNH\ with symmetry $Pnma$. Fe (green) is surrounded by five Cl ions (green) and one H$_2$O molecule (O is shown in red, H in yellow). Thick blue lines are OH bonds, thin blue lines mark hydrogen bonds between FeCl$_5$(H$_2$O) octahedra forming zigzag chains along \bvec, with Fe-Fe distances of 6.47\,\AA. However, the distances to 8 Fe ions of neighboring chains are only slightly larger ($\le$7.04\,\AA). The NH$_4^+$ groups (magenta tetrahedra) are disordered as is modeled by superposing an NH$_4$ tetrahedron with its inversion with respect to the N position, with H drawn in orange and yellow, respectively.\label{struct}}
\end{figure}
Erythrosiderites are another class of materials in which multiferroicity was found \cite{Ackermann2013,Ackermann2014}. In these compounds with $A_2$[Fe$X_5$(H$_2$O)], where $A$ stands for an alkali-metal or an ammonium ion and $X$ for a halide ion, the members are structurally similar, mostly crystallizing in $Pnma$ structure \cite{Bellanca1948, Carlin1977, Figgis1978, OConnor1979, Greedan1980, Puertolas1985, Frohlich2018}, see Fig.~\ref{struct}. Erythrosiderites consist of isolated complex groups of [FeCl$_5$(H$_2$O)]$^{2-}$, where the magnetic iron(III) is six-fold octahedrally coordinated, and of alkali-metal ions or NH$_4^+$ groups. Hydrogen bonds via O-H-Cl interconnect neighboring [FeCl$_5$(H$_2$O)]$^{2-}$ octahedra along the \bvec\ axis and stabilize the structure.
Interestingly, only one member of this group, \pureNH, shows multiferroic order  \cite{Ackermann2013}.
In a two-step magnetic phase transition the ammonium compound first orders antiferromagnetically around \SI{7.2}{\kelvin} \cite{McElearney1978}. In the analogous deuterated material, a collinear sinusoidal spin structure was reported for this phase \cite{Tian2016a, Rodriguez-Velamazan2018}.
At the second transition at \SI{6.7}{\kelvin} the spins reorient and an electric polarization arises \cite{Ackermann2013}. The low-temperature structure was identified to possess a cycloidal order, which causes electric polarization via the inverse Dzyaloshinskii-Moriya interaction \cite{Rodriguez-Velamazan2015, Dzyaloshinsky1958}.
By applying a magnetic field, the polarization can be turned by \ang{90}, which can be explained by a spin-flop transition at $H_{\text{SF}}$ and thus a change of the cycloidal spin structure towards a structure with quasi-collinear antiferromagnetic moments perpendicular to \avec\ \cite{Rodriguez-Velamazan2017}. This transition shows strong magnetoelastic coupling \cite{Ackermann2013, Tian2018} and in the high-field regime the polarization can be explained with the \textit{p}-\textit{d} hybridization mechanism \cite{Rodriguez-Velamazan2017}. Magnetic saturation is reached around \SI{30}{\tesla} \cite{Clune2019}.
\pureK\ exhibits collinear antiferromagnetic order below $T_\text{N}=$ \SI{14.3}{\kelvin} \cite{Ackermann2014, Campo2008}. In this phase, applied magnetic fields induce a polarization via the linear magnetoelectric effect $P_i = \alpha_{ij} \mu_0 H_j$ \cite{Ackermann2014}.
Furthermore, for the multiferroic ammonium compound a high-temperature structural phase transition ($T_\text{S}=\SI{79}{\kelvin}$) is reported, which was attributed to an ordering of hydrogen atoms of the ammonium groups \cite{Ackermann2013}. An analogous transition thus cannot occur in the alkali-metal erythrosiderites and was consequently not observed there.

In this paper, we study the influence of chemical substitution on the low-temperature ordered phases in the substitution series \NHK\ with three different potassium concentrations of $x$ = 0.06, 0.15, and 0.45. This substitution systematically tunes the system from a cycloidal spin structure with a multiferroic ground state towards a collinear antiferromagnet with linear magnetoelectric coupling. Most interestingly, the high-temperature structural transition in \NHK\ is found to be accompanied by a significant change of electric polarization in all partially substituted compounds, which opens a new possibility to control antiferromagnetic domains. Usually, antiferromagnetic antiphase domains cannot be controlled by an external magnetic field. Only for a linear magnetoelectric material and by combining magnetic and electric fields one may drive the antiferromagnetic domains \cite{Brown1998,Baum2013,Kim2018}. In \NHK\, the electric polarization induced by the structural phase transition can take the role of the electric field. Thus, it becomes possible to drive antiferromagnetic domains just by an external magnetic field.

\section{Experimental}
Single crystals of \NHK\ were grown from aqueous solutions. Both end members of this series, \pureNH\ and \pureK\ show incongruent solubility and have to be grown with an excess of FeCl$_3$ and a surplus of HCl. This was also applied for the growth of the mixed crystals. Here, a molar ratio of the solute constituents, FeCl$_3$ and [(1-$x$) NH$_4$Cl + $x$ KCl], of $2 : 1$ was used. Crystal growth was performed by slow temperature reduction between \SI{40}{\degreeCelsius} and \SI{36}{\degreeCelsius} over a period of six months using three solutions with $x \approx 0.067$, $x \approx 0.15$, and $x \approx 0.5$. Large single crystals with dimensions of about \SI{1}{\centi\meter\cubed} were obtained for all three compositions. The lattice constants of these mixed crystals and of the corresponding pure NH$_4$- and K-based end members were obtained via refinement of X-ray powder diffraction data. Assuming the validity of Vegards's rule, i.e., a linear variation of the lattice constants with $x$, actual potassium contents of $x \approx 0.06$, 0.15, and 0.45 are derived for the mixed crystals.

The magnetic susceptibility measurements were performed using a commercial SQUID magnetometer (MPMS, Quantum Design) from \SIrange{2}{300}{\kelvin} in a magnetic field of \SI{0.1}{\tesla}. The temperature and magnetic-field dependencies of the electric polarization were determined by time integration of the induced currents, which were measured by an electrometer (Keithley 6517). Typically, a static electric poling field of about \SI{200}{\volt\per\milli\meter} was applied to ensure single-domain states upon cooling, which, however, has no effect in these samples, as will be seen below. The temperature was changed with various rates between 0.5 and \SI{3}{\kelvin\per\minute}. For the magnetocurrent measurements, the samples were cooled with an electric poling field to various constant temperatures and complete magnetic-field loops (\SI{7}{\tesla} $\rightarrow$ \SI{-7}{\tesla} $\rightarrow$ \SI{7}{\tesla}) were performed with field-sweep rates of \SI{1}{\tesla\per\minute} keeping the poling field constant.
The specific heat was measured between \SI{2}{\kelvin} and \SI{300}{\kelvin} in a commercial calorimeter (PPMS, Quantum Design) by using the thermal relaxation-time method.
Terahertz transmittance data were measured on \pureNH\ using a phase-sensitive continuous-wave THz spectrometer \cite{Roggenbuck2010, Roggenbuck2012} equipped with a linear polarizer and a He bath cryostat.

The crystal structure was investigated on a \ND\ single crystal of ($40\pm2$)\,mg by neutron diffraction using the diffractometer HEiDi \cite{Meven2015} at the MLZ. The crystal was mounted with its \cvec\ axis parallel to the $\phi$ axis of the diffractometer and cooled with a closed-cycle refrigerator. Wavelengths of 1.170~\AA\ and 0.793~\AA\ were obtained from a Ge $ (3\,1\,1) $ resp.~Ge $ (4\,2\,2) $  monochromator in combination with an Er filter. Data were taken with a single detector.

\section{Results and discussion}
\subsection{Structural phase transition}
\begin{figure}
	\includegraphics[width=0.49\textwidth]{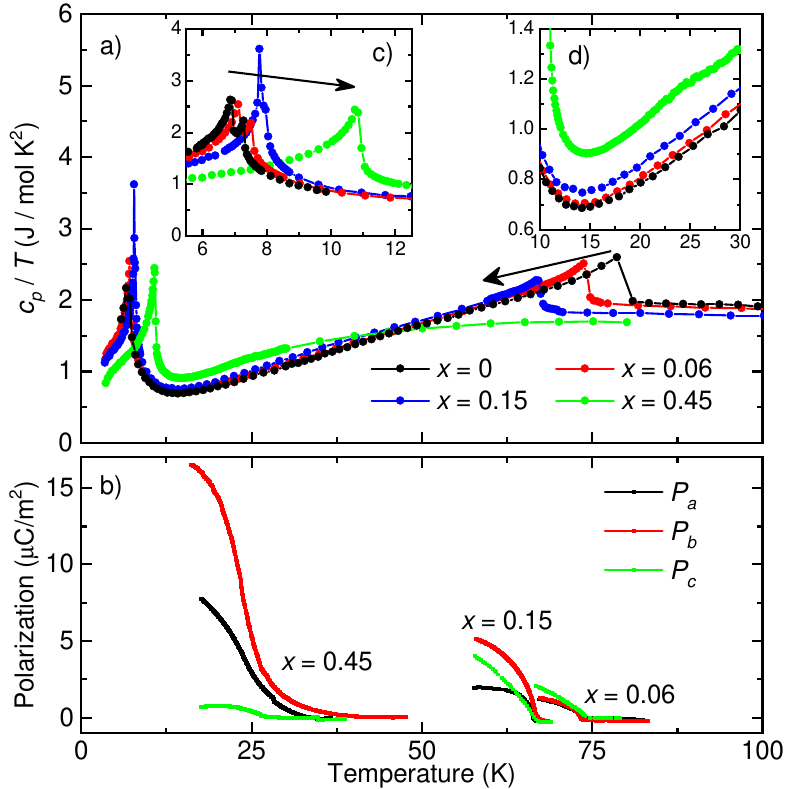}
	\caption{a) Specific heat of the substitution series \NHK. Inset c) resolves the low-temperature transitions and inset d) enlarges the region around the broadened high-temperature transition for $x$ = 0.45 around \SI{24}{\kelvin}. b) Temperature dependence of the polarization components \Pola, \Polb, and \Polc\ for all different potassium concentrations. The pure \pureNH\ does not show a measurable polarization at $T_\text{S}$.}
	\label{spec_heat}
\end{figure}
Figure \ref{spec_heat}a) shows the specific-heat data that reveal several transitions. The structural transition at $T_\text{S} = $ \SI{79}{\kelvin} of pure \pureNH\ systematically shifts to lower temperatures with increasing potassium content ($T_\text{S} (x=0.06) = $ \SI{72}{\kelvin}, $T_\text{S} (x=0.15) = $ \SI{66}{\kelvin}).  For $x=0.45$ the structural transition becomes very broad and occurs around \SI{24}{\kelvin}, see Fig.~\ref{spec_heat}d). In addition, there are two magnetic transitions at $T_\text{N} =$ \SI{7.25}{\kelvin} for the onset of magnetic ordering and at $T_\text{C} =$ \SI{6.8}{\kelvin}, where \pureNH\ becomes ferroelectric, see inset c). For the smallest substitution $x=0.06$ (red) there are still two clearly separated peaks at $T_\text{N}=$ \SI{7.5}{\kelvin} and $T_\text{C}=$ \SI{7}{\kelvin} meaning that both magnetic transitions shift to higher temperature with increasing $x$.
For the higher substitution levels, there is only one anomaly indicating a single second-order phase transition with enhanced transition temperatures $T_\text{N}=$ \SI{7.9}{\kelvin} and \SI{10.8}{\kelvin} for $x=$ 0.15 and 0.45, respectively. The change from two separate transitions to a single transition indicates a change in dielectric and magnetic properties, from spin-spiral multiferroic order for $x\leq 0.06$ to an easy-axis linear magnetoelectric ground state for $x \geq 0.15$, in agreement with the respective ground states of the corresponding pure end members with $x=0$ and 1 \cite{Ackermann2013,Ackermann2014}.

\begin{figure}
	\includegraphics[width=8cm]{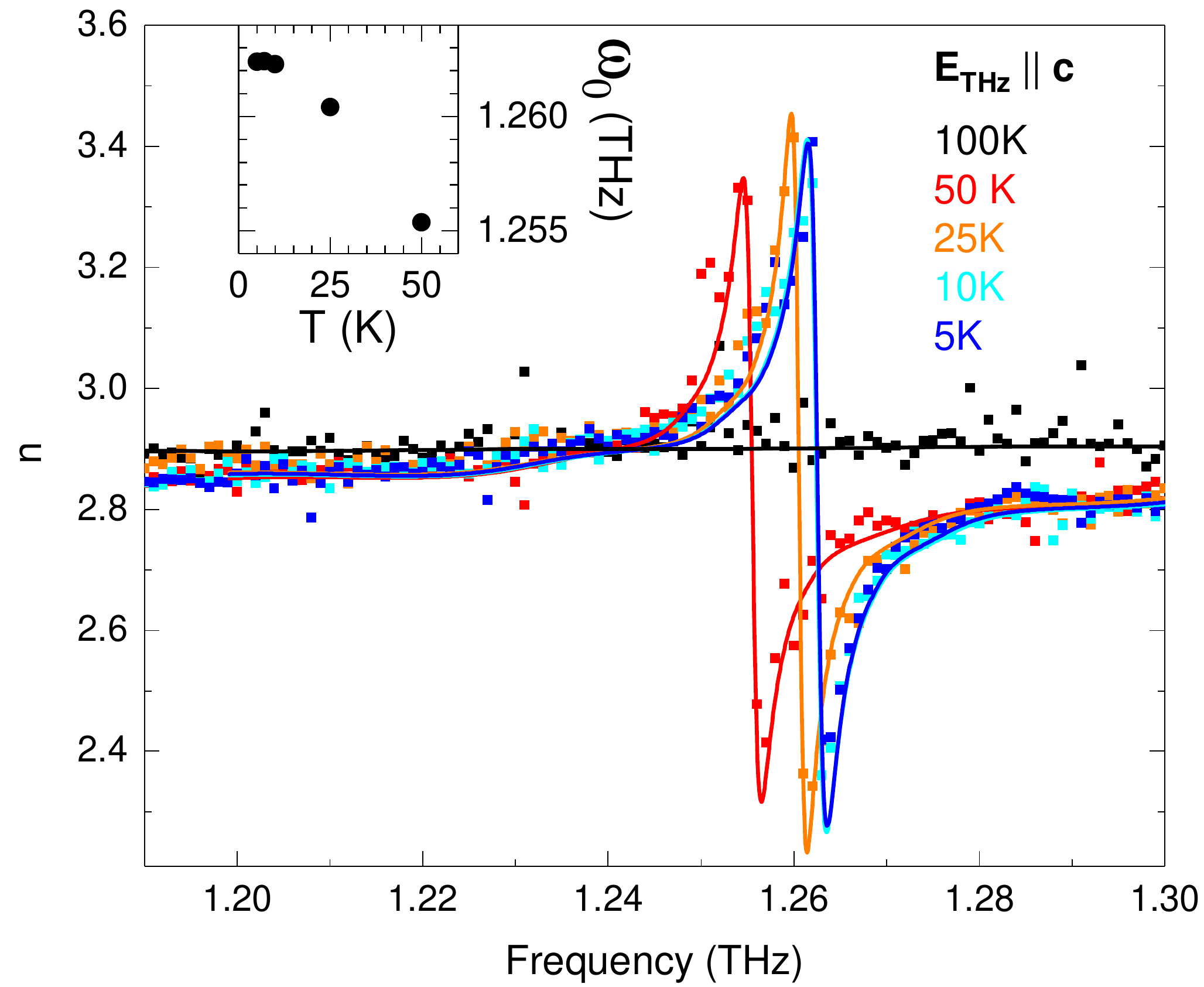}
	\caption{\label{fig:THz}Refractive index of \pureNH\ for polarization $\mathbf{E}\parallel$\cvec.
		Below $T_\text{S}$, we observe a new phonon mode at about 1.26\,THz.
		Symbols show the experimental results, solid lines depict fits using a Lorentz oscillator.
		The inset resolves the temperature dependence of the eigenfrequency of the phonon mode.}
\end{figure}

The high-temperature structural transition around \SI{79}{\kelvin} was proposed to be an ordering of the NH$_4^+$ groups because it is invisible for X-ray diffraction \cite{Ackermann2013,Rodriguez-Velamazan2015}. This scenario naturally explains the decrease of $T_\text{S}$ with increasing $x$, i.e., decreasing NH$_4^+$ concentration, as well as the strong broadening of the corresponding anomaly for $x=0.45,$ where the high degree of substitution prevents the occurrence of a sharp structural phase transition.
Pure \pureNH\ does not show any measurable polarization at $T_\text{S}$ and a high-temperature polarization neither is observed in \pureK , which does not show the structural phase transition. In contrast, for partially substituted \NHK\ a sizable polarization evolves below $T_\text{S}$ and its magnitude systematically increases with increasing $x$ for \Pola\ and \Polb, whereas \Polc\ shows no clear trend within the experimental resolution, see Fig.~\ref{spec_heat}b).
For comparison, the spontaneous polarization in the multiferroic phase of \pureNH\ amounts to approximately \SI{2}{\micro\coulomb\per\meter\squared}, which is even smaller than the absolute value of the high-temperature polarization for $x=0.06$. Surprisingly, the polarization at $T_\text{S}$ is not affected by electric fields up to \SI{200}{\volt\per\milli\meter}. This means that the partially substituted materials are pyroelectric and have a polar structure at least below $T_\text{S}$. We note that these pyrocurrent measurements are only sensitive to polarization changes instead of the absolute polarization.

\begin{figure}
	\includegraphics[width=0.49\textwidth]{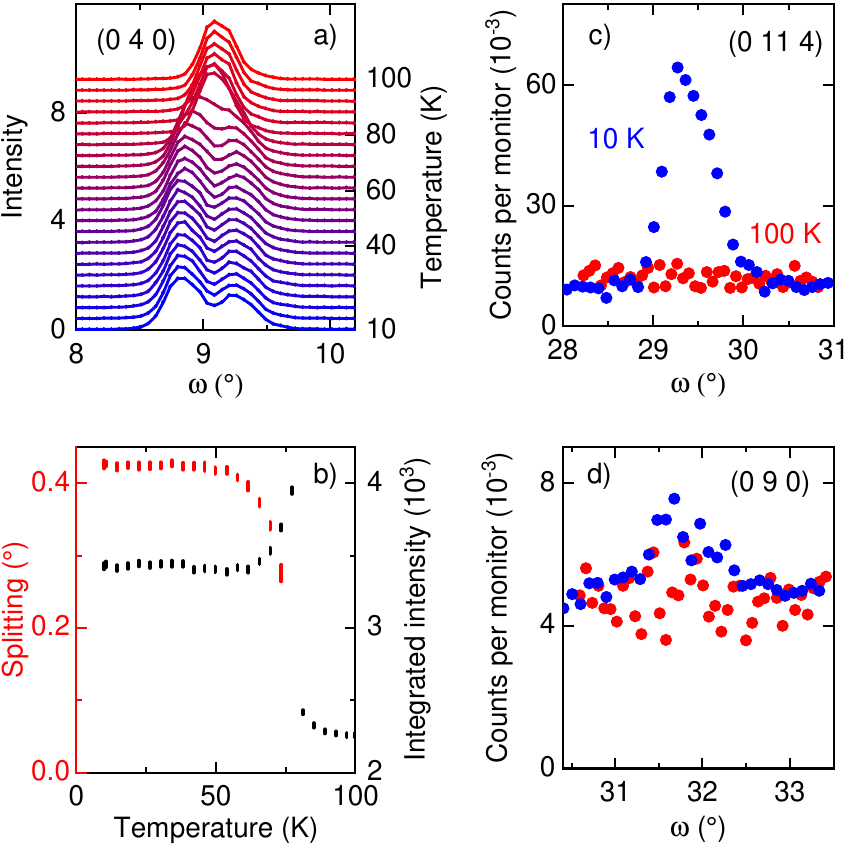}
	\caption{\label{fig:neutrons} Neutron diffraction results on the structural phase transition in \ND .
		Panel a) shows the temperature dependence of the (0 4 0) $\omega$-scan profile, that exhibits a splitting in the
		low-temperature phase due to the appearance of monoclinic domains. Panel b) shows the splitting of the (0 4 0) profiles and
		the integrated intensity. Panel c) and d) show the $\omega$-scans obtained above and below the phase transiton
		for (0 11 4) and (0 9 0), respectively. Note that both reflections are forbidden in the high-temperature symmetry.}
\end{figure}
\begin{figure}
	\flushright\includegraphics[scale=0.3]{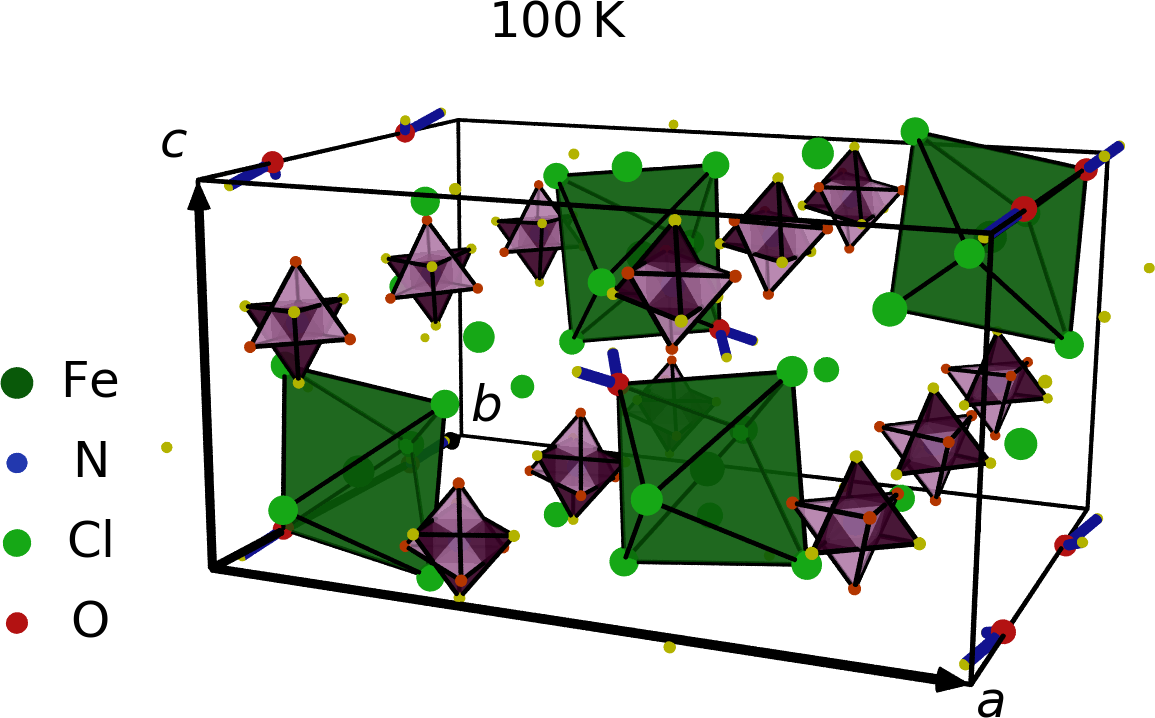}
	\flushright\includegraphics[scale=0.3]{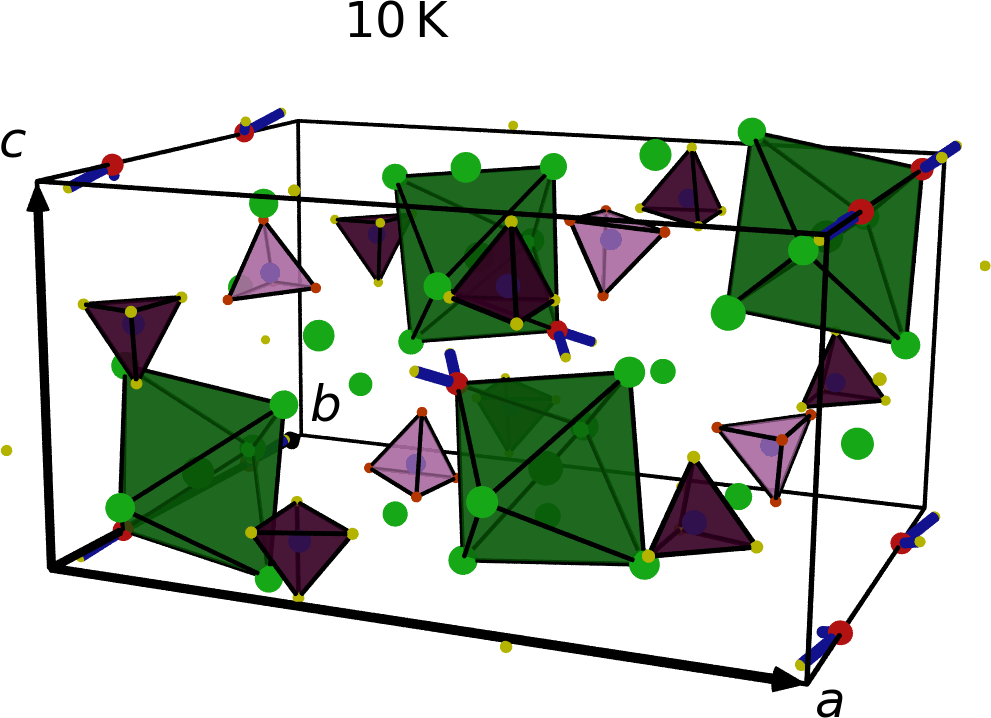}
	\caption{\label{fig:disorder}Crystal structure of $ \mathrm{(ND_4)_2[FeCl_5(D_2O)]} $ at $ 100\,\mathrm{K} $ (top) and $ 10\,\mathrm{K} $ (bottom). $ \mathrm{ND_4^+} $ tetrahedra of the same shading are symmetry equivalent.}
\end{figure}

Clear fingerprints of the structural phase transition are also observed in the phonon spectra.
We studied \pureNH\ by phase-sensitive THz spectroscopy in transmittance geometry
in the frequency range from 0.1 to 1.4\,THz.
Data were collected on freshly polished surfaces. The electric field was either parallel to the \bvec\ or the \cvec\ axis.
Above $T_\text{S}$ = 79\,K, the data reveal insulating behavior and a nearly frequency-independent
refractive index $n(\omega)\approx 3$.
The small observed dispersion in $n(\omega)$ can be described by assuming a phonon mode with an eigenfrequency
above 2\,THz.
However, below $T_\text{S}$ a feature with an eigenfrequency of about 1.26\,THz ($\approx$ 5.2\,meV) appears
in the studied frequency range, see Fig.~\ref{fig:THz}, which signals the structural phase transition. This new mode may either be attributed to a change of selection rules at the phase transition or to a strong shift of its eigenfrequency.
The eigenfrequency shifts by about 0.6\,\% between 5\,K and 50\,K, see inset of Fig.~\ref{fig:THz}. 
In many ammonium compounds an order-disorder transition of the NH$_4^+$ group is known, e.g., in the large family of alums \cite{RamaRao1978}.
In an ordered structure with preferred NH$_4$ orientation, the ammonium ions may show libration, i.e., torsional oscillation around the equilibrium orientation. Librational frequencies vary strongly for different compounds \cite{Smith1985}. Excitation energies of, e.g., 5.9\,meV and 4\,meV were observed in neutron scattering on (NH$_4$)$_2$PdCl$_6$ \cite{Prager1984} and K$_{1-x}$(NH$_4$)$_x$SCN \cite{Smirnov2009}, respectively, suggesting a common origin with the 5.2\,meV mode in \pureNH. A disordered phase, in contrast, may show a much lower potential barrier between different orientations and thus a very different excitation energy.

In order to examine the structural phase transition, \textsc{}the crystal structure of \ND\ was studied in the paramagnetic state at $ 100\,\mathrm{K} $ (above $ T_{\text{S}} $) and at $ 10\,\mathrm{K} $ (below $ T_{\text{S}} $) by single-crystal neutron diffraction. The results of temperature-dependent studies are presented in Fig.~\ref{fig:neutrons}. The (0 4 0) Bragg reflection splits
below the structural phase transition indicating a monoclinic distortion. The temperature dependence of this monoclinic distortion well agrees with an order-disorder character
of the transition. 
There is a clear peak in the integrated intensity of this reflection just at $T_\text{S}$, which points
to the impact of the ferroelastic domains on the extinction. Close to the phase transition the domain structure severely reduces the extinction conditions and thereby enhances the scattering signal, and also the fixed domain structure at lower temperature corresponds to a lower crystal quality, less extinction and thus higher intensities. The glide plane $n$ in space group
$Pnma$ implies a special selection rule that $k+l$ must be even for $(0\,k\,l)$ reflections. The observation of reflections at low temperature that violate these special selection rules, see
Fig.~\ref{fig:neutrons} c) and d), unambiguously shows that this glide plane is broken below $T_\text{S}$. There are, however, no superstructure reflections that can be associated
with a loss of translation symmetry. The structural distortion must be described by a $\Gamma$ mode \cite{Stokes1989}.

Full data sets of Bragg reflection intensities were collected at $ 100\,\mathrm{K} $ (in total $ 3\,335 $ reflections) and  at $ 10\,\mathrm{K} $ ($ 5\,204 $ reflections). The data collected at $ 100\,\mathrm{K} $ ($ 10\,\mathrm{K}) $ yield an internal $ R $ value of $ 2.02\,\% $ ($ 1.53\,\% $) when the data are reduced according to space group $ Pnma $ ($ P2_1/a $). The refinement with the $ 100\,\mathrm{K} $ data in space group $ Pnma $ yields two different orientations of ND$_4^+$ tetrahedra with almost equal partial occupation of $ 56.13(16)\,\% $ and $ 43.87(16)\,\% $, respectively. All refined structure models contain isotropic atomic displacement parameters (ADP) for the $ \mathrm{Fe} $ atoms and anisotropic ADPs for all other atoms. Nearly opposing $ \mathrm{D} $ atoms of the two possible orientations are constrained to have the same ADPs and a total occupancy of $ 1 $. The $ \mathrm{D} $ atoms within one tetrahedron are constrained to have the same occupancy. The refinement of the model with two independent tetrahedra describes the data well with satisfactory reliability parameters, given in table \ref{tab:table_positions}.
In order to further simplify the model and to reduce the number of free parameters, we constrain the second $\mathrm{D}_4$ tetrahedron to arise from the first one just by inversion with respect to the central $\mathrm{N}$ atom, which results in an only slightly worsened description, ($\mathrm{w}R(\mathrm{all})=6.09\,\%$). This disordered structure is illustrated in Figs. 1 and 5, and
the corresponding structural parameters together with the $ R $ values are given in Table \ref{tab:table_positions}.

\begin{table}
	\begin{tabular}{c c c c c}\hline
		\multicolumn{2}{c}{$ 100\,\mathrm{K}, Pnma $} & \multicolumn{3}{c}{$ a = 13.579, b = 9.922, c = 6.947\,\mathrm{\AA} $}\\
		& \multicolumn{2}{c}{$ R(\mathrm{obs}) = 4.86\,\% $} & \multicolumn{2}{c}{$ \mathrm{w}R(\mathrm{obs}) = 5.95\,\% $} \\
		& \multicolumn{2}{c}{$ R(\mathrm{all}) = 7.97\,\% $} & \multicolumn{2}{c}{$ \mathrm{w}R(\mathrm{all}) = 6.09\,\% $} \\\hline
		& $ x $ & $ y $ & $ z $ & $ U_{\mathrm{iso}} $ \\\hline
		Fe1 & $ 0.11807(8) $ & $ 1/4 $ & $ 0.1900(2) $ & $ 7.1(3) $\\
		Cl1 & $ 0.10580(6) $ & $ 0.00997(8) $ & $ 0.1763(1) $ & $ 10.7(3) $\\
		Cl2 & $ 0.00679(9) $ & $ 1/4 $ & $ 0.4576(2) $ & $ 11.4(4) $\\
		Cl3 & $ 0.25069(9) $ & $ 1/4 $ & $ 0.4004(2) $ & $ 9.5(4) $\\
		Cl4 & $ 0.22553(9) $ & $ 1/4 $ & $ -0.0756(2) $ & $ 11.2(4) $\\
		O1 & $ -0.0028(2) $ & $ 1/4 $ & $ 0.0000(4) $ & $ 14.9(6) $\\
		D1 & $ -0.0342(1) $ & $ 0.1712(1) $ & $ -0.0502(3) $ & $ 25.2(5) $\\
		N1 & $ 0.13865(6) $ & $ 0.00028(9) $ & $ 0.6653(2) $ & $ 13.5(3) $\\
		D2 & $ 0.1356(2) $ & $ 0.0079(4) $ & $ 0.8133(5) $ & $ 46(1) $\\
		D3 & $ 0.1067(3) $ & $ -0.0817(3) $ & $ 0.6225(6) $ & $ 44(1) $\\
		D4 & $ 0.2120(2) $ & $ 0.0040(4) $ & $ 0.6352(5) $ & $ 40(1) $\\
		D5 & $ 0.1059(2) $ & $ 0.0835(3) $ & $ 0.6115(6) $ & $ 40(1) $\\
		\hline
		~\\\hline
		\multicolumn{2}{c}{$ 10\,\mathrm{K}, P2_1/a $} & \multicolumn{3}{l}{$ a = 13.553, b = 9.960, c = 6.925\,\mathrm{\AA} $}\\
		& & \multicolumn{3}{l}{$ \gamma = 90.2^{\circ} $} \\
		& \multicolumn{2}{c}{$ R(\mathrm{obs}) = 3.38\,\% $} & \multicolumn{2}{c}{$ \mathrm{w}R(\mathrm{obs}) = 4.30\,\% $} \\
		& \multicolumn{2}{c}{$ R(\mathrm{all}) = 5.39\,\% $} & \multicolumn{2}{c}{$ \mathrm{w}R(\mathrm{all}) = 4.41\,\% $} \\\hline
		& $ x $ & $ y $ & $ z $ & $ U_{\mathrm{iso}} $ \\\hline
		Fe1 & $ 0.11923(6) $ & $ 0.2499(1) $ & $ 0.1875(1) $ & $ 2.8(2) $\\
		Cl1\_1 & $ 0.10315(7) $ & $ 0.0103(1) $ & $ 0.1825(1) $ & $ 4.7(3) $\\
		Cl1\_2 & $ 0.10981(7) $ & $ 0.4897(1) $ & $ 0.1667(1) $ & $ 4.5(2) $\\
		Cl2 & $ 0.00651(6) $ & $ 0.2551(1) $ & $ 0.4567(1) $ & $ 4.6(2) $\\
		Cl3 & $ 0.25158(6) $ & $ 0.2489(1) $ & $ 0.4001(1) $ & $ 4.6(2) $\\
		Cl4 & $ 0.22700(6) $ & $ 0.2451(1) $ & $ -0.0768(1) $ & $ 4.9(2) $\\
		O1 & $ -0.0017(1) $ & $ 0.2492(2) $ & $ -0.0033(2) $ & $ 7.1(3) $\\
		D1\_1 & $ -0.0346(1) $ & $ 0.1705(2) $ & $ -0.0530(3) $ & $ 20.1(6) $\\
		D1\_2 & $ -0.0353(1) $ & $ 0.3280(2) $ & $ -0.0500(3) $ & $ 22.0(6) $\\
		N1\_1 & $ 0.13640(7) $ & $ -0.0028(1) $ & $ 0.6643(2) $ & $ 7.1(3) $\\
		N1\_2 & $ 0.13987(7) $ & $ 0.4949(1) $ & $ 0.6700(2) $ & $ 7.4(3) $\\
		D2\_2 & $ 0.1315(1) $ & $ 0.4918(2) $ & $ 0.8162(3) $ & $ 22.5(5) $\\
		D3\_2 & $ 0.1110(2) $ & $ 0.5825(2) $ & $ 0.6176(3) $ & $ 27.2(6) $\\
		D4\_2 & $ 0.2132(1) $ & $ 0.4888(2) $ & $ 0.6400(3) $ & $ 26.1(6) $\\
		D5\_2 & $ 0.1031(1) $ & $ 0.4161(2) $ & $ 0.6086(3) $ & $ 22.5(6) $\\
		D6\_1 & $ 0.1407(1) $ & $ -0.0077(3) $ & $ 0.5172(3) $ & $ 26.9(6) $\\
		D7\_1 & $ 0.1702(2) $ & $ 0.0829(2) $ & $ 0.7138(3) $ & $ 25.6(6) $\\
		D8\_1 & $ 0.0637(1) $ & $ -0.0019(2) $ & $ 0.7017(3) $ & $ 21.1(5) $\\
		D9\_1 & $ 0.1702(1) $ & $ -0.0852(2) $ & $ 0.7227(3) $ & $ 24.1(6) $\\
		\hline
	\end{tabular}
	\caption{\label{tab:table_positions}\ND: Results of the structural refinements with the neutron-diffraction data sets obtained at 100\,K and 10\,K in space groups $Pnma$ and $P2_1/a$, respectively. Lattice parameters were taken from reference \onlinecite{Ackermann2013} (besides $\gamma$ deduced from the splitting of the ($0\,4\,0$) reflection) and anisotropic atomic displacement parameters
		are given in the Appendix.}
\end{table}

Starting from the high-temperature $Pnma$ phase, we first consider possible symmetry reductions that result in a centrosymmetric structure and that do not break translation symmetry. The modes $ \Gamma_2^+ $, $ \Gamma_3^+ $, and $ \Gamma_4^+ $ lead to space groups $P2_1/a$, $P2_1/n$, and $P2_1/m$, respectively \cite{Stokes1989, footnote}. They can be distinguished by their selection rules. We can rule out space group $ P2_1/n $ because we observed reflections of the form $ (0\,k\,0) $ with odd $ k $ and $ (0\,k\,l) $ with odd $ k + l $, see Fig.~\ref{fig:neutrons}c) and d). Space group $P2_1/a$ (we use a non-conventional setting in order to allow for direct comparison with the high-temperature $Pnma$ structure) is in accordance with all observed, respectively not observed reflections. Space group $ P2_1/m $ allows reflections of the form $ (h\,k\,0) $ with odd $ h $, which are not observed, but this does not rule out this space group, because the reflections might be below the detection level.
Structural refinements were thus carried out in space groups $ Pnma $, $ P2_1/a $, and $ P2_1/m $ using the software JANA2006 \cite{Petricek2014}. The $ R $ values \cite{Karplus2012} unambiguously show that the neutron diffraction data at $ 10\,\mathrm{K} $ are best described in space group $ P2_1/a $. Among the symmetry reductions resulting in a non-centrosymmetric low-temperature structure \cite{Stokes1989} only space group $Pmc2_1$ is in accordance with the observed superstructure reflections, but a satisfactory
description of the data cannot be achieved in this symmetry. A non-centrosymmetric structure can also be obtained by further reducing the symmetry from $ P 2_1/a $ to
$ P a $, but the refinements of this model give no evidence for such a distortion. We thus conclude that the low-temperature symmetry corresponds to $ P 2_1/a $ in
agreement with the absence of a spontaneous electric polarization at the structural transition in the pure material. This monoclinic symmetry $P2_1/a$ is in agreement with the occurrence of ferroelastic domains of two different orientation states at $T_\text{S}$ and with the orientation of the optical indicatrix within these domains, as it was found in Ref.~\onlinecite{Ackermann_Dis}.

The refinement of the structure model in space group $ P2_1/a $ with the $ 10\,\mathrm{K} $ data  yields a structure, in which each of the 8 tetrahedra within the unit cell strongly prefers one of the two possible orientations ($ 95.3(3)\,\% $ resp.~$ 95.5(2)\,\% $). Due to the low occupation of the minority tetrahedra their structural parameters cannot be reliably refined. We therefore
simplified the structure model by assuming perfect tetrahedron ordering which again only slightly reduces the reliability values ($\mathrm{w}R(\mathrm{all})=4.41\,\%$).
The resulting structure is shown in Fig.~\ref{fig:disorder} and structural parameters together with the $ R $ values are given in Table \ref{tab:table_positions}. There are two independent distorted NH$_4^+$ groups in space group $ P2_1/a $ that are related through the $n$ glide plane in $Pnma$; the centrosymmetry, however, persists so that for each of these independent NH$_4^+$ groups there exists an inverted one.

In conclusion, the neutron diffraction data unambiguously show that the structural phase transition in \ND\ at $ T_{\text{S}} $ results from orientational order of the ND$_4^+$ tetrahedra, in accordance with Ref.~\cite{Rodriguez-Velamazan2015}. These ordered tetrahedra show a libration mode in THz spectroscopy. Although a satisfying description of the Bragg reflection intensities is obtained within the centrosymmetric but monoclinic structure, this solution does not rule out the appearance of a small but finite electric polarization in the structurally distorted phase for the partially K-substituted samples. In the pure compound the dipole moments of the NH$_4^+$ groups cancel so that the total structure is not polar.
However, this balance can be perturbed by the partial substitution with non-polar K$^+$ ions, in particular, if they stay mobile around the structural phase transition.

Since the lattice becomes monoclinic in the distorted structure, there are only two equivalent space diagonals, along which the finite polarization could appear, but the measured polarization, see Fig.~\ref{spec_heat}b) does not follow this explanation, pointing to a more exotic mechanism. The experimentally determined polarization vector is only allowed in a triclinic structure, which is not supported by diffraction data on the substituted compounds. An in-grown potassium concentration gradient could cause a built-in electric field and a corresponding polar structure, as it is proposed for ferroelectric KTiOPO$_4$ \cite{Marvan2005,Dhanaraj2010}. However, pure \pureNH \ is paraelectric. The inversion symmetry
can be broken by partial ordering of the K substitution. This loss of inversion symmetry implies strong pyroelectric effects at the structural phase transition, where the
NH$_4^+$ groups order due to a finite dipole moment in the low-symmetry environment.

\subsection{Low-temperature transitions}
\begin{figure}
	\includegraphics[width=0.49\textwidth]{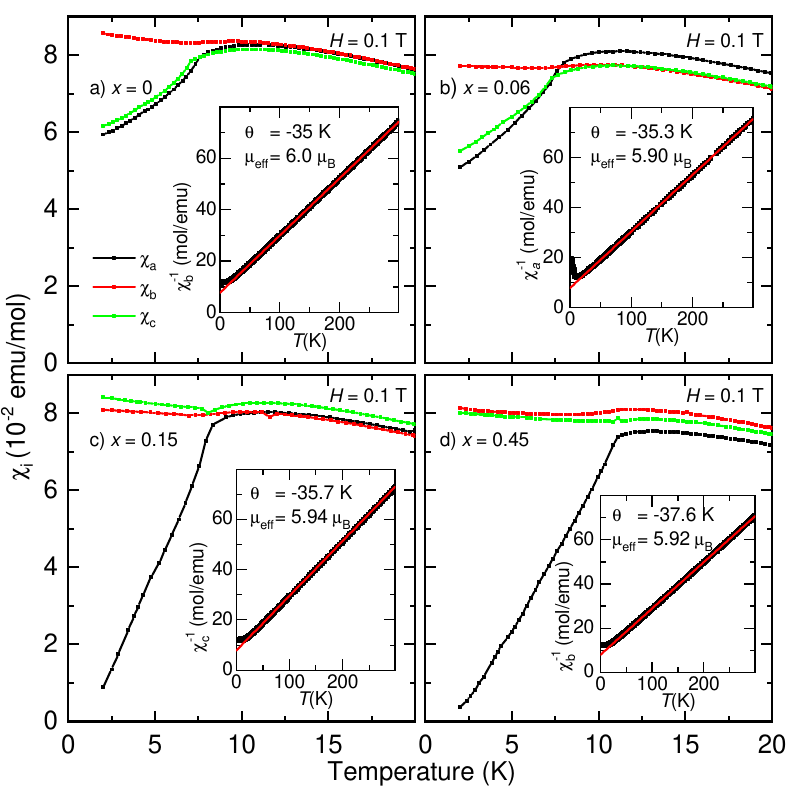}
	\caption{Susceptibility of the substitution series \NHK\ with a) $x=$ 0, b) $x=$ 0.06, c) $x=$ 0.15, d) $x=$ 0.45. The insets show the Curie-Weiss fits.}
	\label{susceptibility}
\end{figure}

In Fig.~\ref{susceptibility}, we present the magnetic susceptibility of the substitution series \NHK. As discussed in Ref.~\onlinecite{Ackermann2013}, the susceptibility of \pureNH\ shows a high-temperature Curie-Weiss behavior and an $ac$ easy-plane anisotropy below $T_\text{N}$ indicating a spin-spiral order. For $x=0.06$, we find an analogous behavior of $\chi(T)$, in particular the evolution of an easy-plane anisotropy indicating that this compound hosts a cycloidal spin texture as well.
For potassium concentrations of 0.15 (panel c) and 0.45 (panel d), both $\chi_b$ and $\chi_c$ remain constant below $T_\text{N}$ of \SI{7.9}{\kelvin} and \SI{10.8}{\kelvin}, respectively. In contrast, $\chi_a$ drops towards zero, which indicates collinear antiferromagnetic order with moments along \avec\ as it is also found in \pureK~\cite{Ackermann2014}. As already known from the specific-heat data (Fig.~\ref{spec_heat}), the N\'{e}el temperature increases with increasing $x$ and the transition
only splits for $x$ = 0 and 0.06. For all substitution levels, the high-temperature susceptibility follows a Curie-Weiss law with nearly the same effective magnetic moment of approximately 5.95 $\mu_\text{B}$ (see insets), which well agrees with the expected value  of 5.92 $\mu_\text{B}$ for $S=5/2$. Additionally, the absolute value of the Curie-Weiss temperature increases slightly with increasing potassium content, from $\theta =$ \SI{-34.9}{\kelvin} to \SI{-37.6}{\kelvin}.
The fact that $\left|\theta\right|>T_\text{N}$ indicates magnetic frustration in all systems. Because $\left|\theta\right|$ increases, the net antiferromagnetic exchange becomes stronger with increasing $x$, but as $T_\text{N}$ increases even faster, the frustration parameter $f=\left|\theta\right|/ T_\text{N}$ is reduced continuously from $f\approx4.8$ in \pureNH\ to 2.8 in \pureK. This suggests that the change of the magnetic order from cycloidal to collinear is directly linked to the reduced magnetic frustration.

\begin{figure}
	\includegraphics[width=0.49\textwidth]{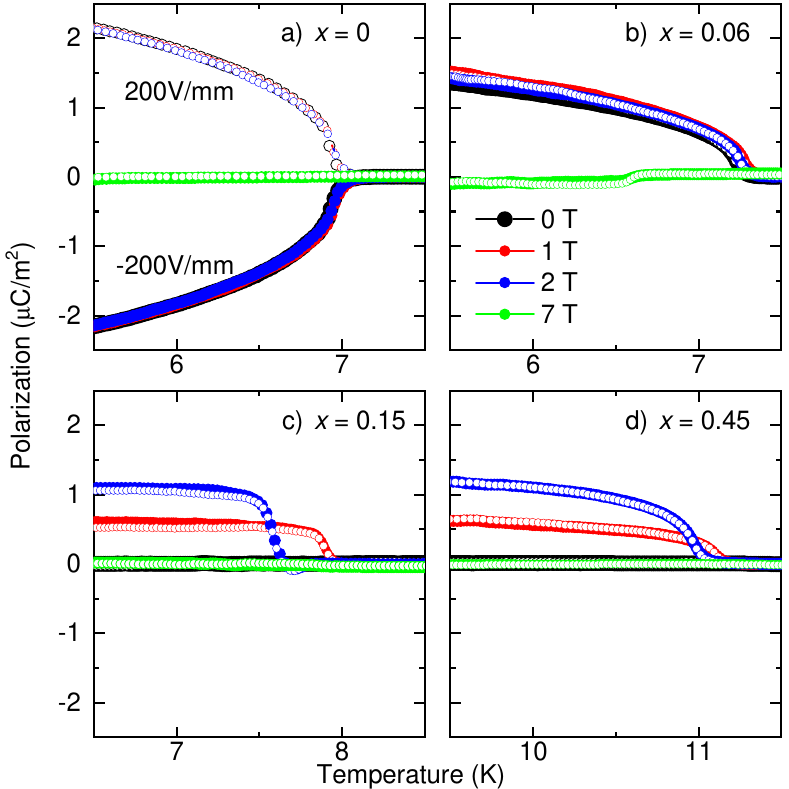}
	\caption{Temperature dependence of the polarization \Pola\ for the substitution series \NHK\ with a) $x=$ 0, b) $x=$ 0.06, c) $x=$ 0.15, d) $x=$ 0.45. Upon cooling, electric poling fields of $\pm$\SI{200}{\volt\per\milli\meter} (open and closed symbols, respectively) were applied and removed at base temperature. The polarization was then obtained by numerical integration of the pyrocurrent measured upon heating.}
	\label{PvT}
\end{figure}

Figure~\ref{PvT} displays the temperature-dependent polarization of the substitution series. Again, we include the data for \pureNH\ from Ref.~\onlinecite{Ackermann2013}, which show a typical type-II multiferroic with a rather small polarization. In low magnetic fields, this polarization is almost constant and mainly points along \avec , but switches towards the \cvec\ direction due to a spin-flop transition occurring around \SI{5}{\tesla} \cite{Ackermann2013}. Panel~b) shows the analogous data for the smallest potassium content, which also exhibits a spontaneous polarization  at zero magnetic field that hardly varies in small fields.
As discussed above, \NHK\ already shows a finite polarization below the structural transition. Therefore, the polarization change occurring at low-temperature is poled by this internal polarization which causes an internal electric field $E_\text{int}=\frac{2}{3\epsilon_0}P\approx\SI{750}{\volt\per\milli\meter}$. Consequently, external poling fields of \SI{200}{\volt\per\milli\meter} cannot switch the direction of the additional polarization.
Thus, we call this phase pseudo-ferroelectric and pseudo-multiferroic. In larger magnetic fields, the additional polarization along \avec\ again almost fully vanishes, because it switches towards the \cvec\ direction, in analogy to the pure \pureNH\ (see below).

For higher substitution levels, $x=0.15$ and 0.45, the magnetoelectric behavior qualitatively changes. As shown in panels c) and d), there no longer is an additional spontaneous polarization in zero magnetic field, but small applied magnetic fields induce an extra electric polarization, whose magnitude first linearly increases with magnetic field and then abruptly vanishes because again the polarization switches towards the \cvec\ direction (see below). This kind of linear magnetoelectric behavior is very similar to that of the pure \pureK\ and other alkaline-metal-based erythrosiderites~\cite{Ackermann2014}. One main difference is, however, that again the polarization cannot be switched by electric poling fields of \SI{200}{\volt\per\milli\meter}, because of the polarization that appears already at $T_\text{S}$ (see Fig.~\ref{spec_heat}b).

\begin{figure}
	\includegraphics[width=0.49\textwidth]{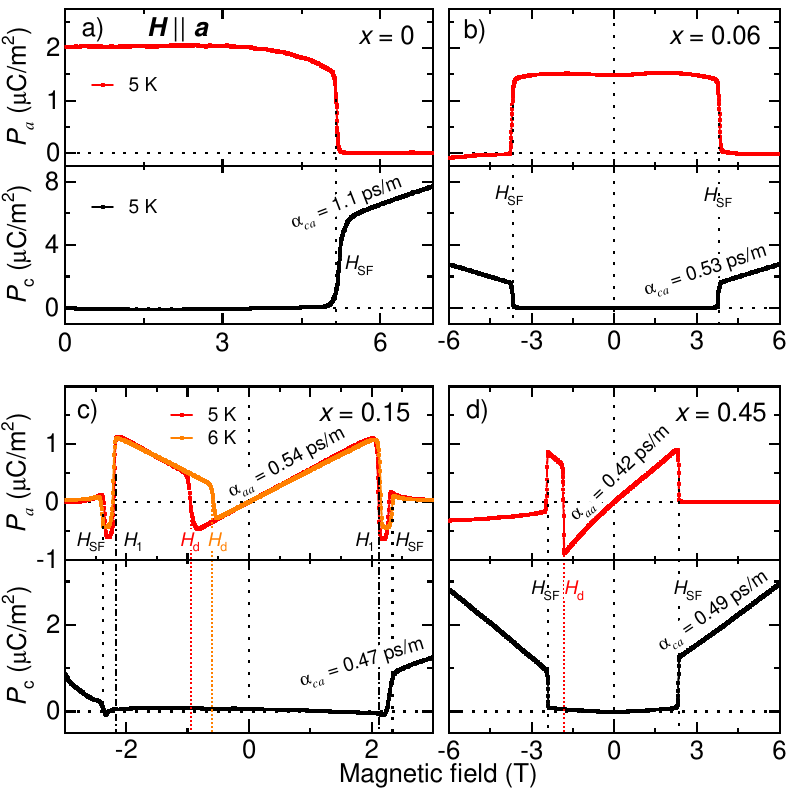}
	\caption{Magnetic-field dependence of the polarization components \Pola\ (red) and \Polc\ (black) for \Hvec$\parallel$\avec\ for the substitution series \NHK\ with a) $x=$ 0, b) $x=$ 0.06, c) $x=$ 0.15, d) $x=$ 0.45. All polarization curves were obtained by numerical integration of the magnetocurrent measured as a function of increasing magnetic field. For $x=0$, electric poling fields of \SI{200}{\volt\per\milli\meter} were applied. $H_\text{SF}$ denotes the spin-flop field, which is accompanied by a polarization switch from the \avec\ to the \cvec\ direction. For $x=0.15$, an intermediate phase with inverted polarization \Pola\ occurs in the field range $H_1<H<H_\text{SF}$. Switching of magnetoelectric domains at $H_\text{d}$ is observed for $x = 0.15$ and $x = 0.45$. The linear magnetoelectric coefficients $\alpha_{ij} = \partial P_i / \partial (\mu_0H_j)$ are also given.}
	\label{P(B)_alle}
\end{figure}
We also determined the magnetic-field dependence of the polarization directly by integrating the measured magnetic-field dependent (magneto-)current. In Fig.~\ref{P(B)_alle}, we present both polarization components \Pola\ and \Polc\ as a function of \Hvec $\parallel$\avec. For $x=0$, panel a) shows an approximately constant polarization along \avec\ up to $\mu_0H\simeq$ \SI{5}{\tesla}, where it switches to \cvec\ and then linearly increases with further increasing magnetic field, i.e., $P_c = \alpha_{ca} \mu_0 H_a$ \cite{Ackermann2013}.
Figure \ref{P(B)_alle}b) displays the polarization components \Pola\ and \Polc\ for $x = 0.06$, which behave similarly to the pure compound. However, there are also clear differences. First of all, the switching field is reduced to \SI{3.5}{\tesla} for $x=0.06$ and also the linear magnetoelectric coefficient of \Polc\ is reduced to $\alpha_{ca}=\partial P_c/\partial (\mu_0H_a)=\SI{0.53}{\pico\second\per\meter}$. As discussed above, for $x=0.06$ the sign of the polarization is not switchable by applying an electric field, in contrast to the pure $x=0$ compound. Above $H_\text{SF}$, the polarization \Polc\ is of the same sign for both magnetic-field directions and cannot be inverted by electric fields up to \SI{200}{\volt\per\milli\meter}, again in contrast to the $x=0$ compound \cite{Ackermann2013}. The reason is that the preexisting internal electric field for $x=0.06$ fixes the direction of \Polc, which implies that opposite linear magnetoelectric domains are realized for positive and negative magnetic field.

For larger $x$, see panels c) and d), the dielectric properties change drastically, because these materials are no longer multiferroic or pseudo-multiferroic. We start the discussion with the $x=0.45$ sample. Above $H_\text{SF}$, this material is again linear magnetoelectric with the sign of \Polc\ remaining unchanged for both magnetic-field directions, in analogy to the $x=0.06$ sample. Below $H_\text{SF}$, the $x=0.45$ sample also shows linear magnetoelectric coupling $P_a = \alpha_{aa} \mu_0H_a$. Remarkably, there is a polarization sign reversal at $H_\text{d}=$ \SI{-1.6}{\tesla} for \SI{5}{\kelvin}, and below this flop the slope $\partial P_a/\partial (\mu_0 H)$ is inverted as well. In the linear magnetoelectric phase, there are in general two possible domains, which are usually switchable by applying an external electric field (in addition to $H$). In the present case, however, the high-temperature polarization, see Fig.~\ref{spec_heat}b), provides an intrinsic electric field that poles the magnetoelectric domains. Due to this combination, it is possible to switch between linear magnetoelectric domains with a magnetic field even without any external electric field.

\begin{figure}
	\includegraphics[width=0.49\textwidth]{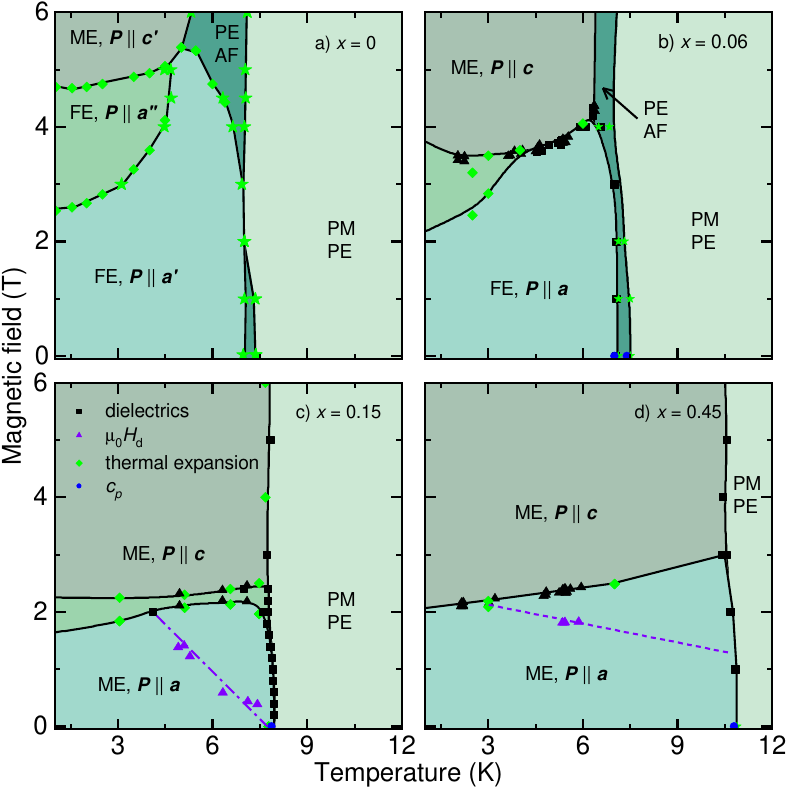}
	\caption{Phase diagrams of the substitution series \NHK\ with a) $x=$ 0, b) $x=$ 0.06, c) $x=$ 0.15, d) $x=$ 0.45 for \Hvec$\parallel$\avec. Phase boundaries are based on anomalies in thermal expansion and magnetostriction (not discussed here), pyrocurrent and magnetocurrent (dielectrics), and specific heat measurements. The points of $\mu_0H_\text{d}$ stem from the magnetic-field dependent anti-phase domain switching discussed for Fig.~\ref{P(B)_alle}. The phases labeled ME show the linear magnetoelectric effect, FE phases are (pseudo-)ferro\-elec\-tric, PM = paramagnetic, AF = antiferromagnetic, PE = paraelectric.}
	\label{Phasendiagramm_alle}
\end{figure}
For $x=$ 0.15, see panel c), the material is still linear magnetoelectric as we find a linear polarization \Pola\ for small magnetic fields. The domain switching field $H_\text{d}$ decreases with increasing temperature, indicating activated behavior. Between $|H_1|=$ \SI{2.2}{\tesla} and $|H_{\text{SF}}|=$ \SI{2.4}{\tesla}, an intermediate phase occurs with a negative and essentially constant polarization. Above $H_{\text{SF}}$, the polarization switches to the \cvec\ direction, as for $x=0.45$, and grows linearly in magnetic fields. The intermediate regime might be a precursor of the ferroelectric intermediate phase known for \pureNH.

To summarize these findings, the spin-flop phases above $H_\text{SF}$ are linear magnetoelectric for all samples. Interestingly, all mixed samples even quantitatively show very similar magnetoelectric coefficients below and above the spin-flop transition with $\alpha_{ca} \approx \alpha_{aa} \approx\SI{0.5}{\pico\second\per\meter}$, whereas the $x=0$ parent compound has a larger $\alpha_{ca}=\SI{1.1}{\pico\second\per\meter}$. Concerning the switching of magnetoelectric domains, it is not possible to switch the polarization above $H_\text{SF}$ by an external electric field of up to \SI{200}{\volt\per\milli\meter}. This means that opposite antiferromagnetic domains are realized for larger positive and negative magnetic fields, i.e., for $\pm H>|H_\text{SF}|$, in order to align the induced magnetoelectric polarization with \Polc\ that is present already above $T_\text{N}$. In particular, these are single-domain states even in zero electric field, whereas usual linear magnetoelectrics with spin-flop transitions, e.g., \pureNH\ \cite{Ackermann2013} require the application of both a magnetic and an electric field to reach a single-domain state above $H_\text{SF}$.
\begin{figure}
	\includegraphics[width=0.49\textwidth]{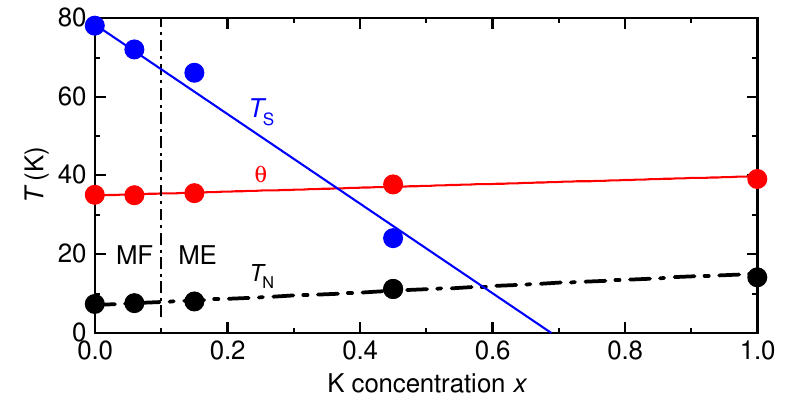}
	\caption{Potassium concentration dependence of the transition temperatures $T_\text{N}$, $T_\text{S}$, and the magnetic Curie-Weiss temperature $\theta$ for \NHK.}
	\label{diltion_overview}
\end{figure}

Finally, we present the resulting temperature vs. magnetic-field phase diagrams with \Hvec$\parallel$\avec\ for all investigated substitution levels in Fig.~\ref{Phasendiagramm_alle}. Pure \pureNH\ is a type-II multiferroic, which has been characterized in detail \cite{Ackermann2013,Rodriguez-Velamazan2015,Rodriguez-Velamazan2017,Clune2019}. For $x=0.06$ we still find a two-step magnetic ordering transition ($T_\text{N}=$ \SI{7.25}{\kelvin}, $T_\text{C}=$ \SI{7}{\kelvin}), which is typical for multiferroics with a cycloidal spin texture. Similar to the pure ammonium-based compound, there is an intermediate-field phase around \SI{3}{\tesla} that vanishes above \SI{4}{\kelvin}. The low-field regime shows a pseudo-multiferroic non-zero spontaneous polarization \Pola. The term pseudo-multiferroic is used because the polarization cannot be switched by applying electrical poling fields (of the order of \SI{200}{\volt\per\milli\meter}) because the structure is already polar for $T>T_\text{C}$.

The exact direction of the polarization in the intermediate phase around \SI{3}{\tesla} was not part of this study, but one may suspect that there might be a slight reorientation of the polarization as it is the case in the pure \pureNH\ due to a reorientation of the magnetic moments, see Refs.~\onlinecite{Ackermann2013,Rodriguez-Velamazan2017}. The potassium content of $x=$ 0.15 suppresses the spontaneous polarization and changes the cycloidal spin structure towards a collinear order with a single antiferromagnetic transition at \SI{7.9}{\kelvin}. The material is linear magnetoelectric with the polarization pointing along \avec\ with \Hvec$\parallel$\avec. When performing magnetic-field hysteresis loops, it is possible to switch between the two different linear-magnetoelectric domains. The switching is thermally activated and the switching field strongly increases with decreasing temperature. Below \SI{4}{\kelvin}, switching is no longer possible because there is a transition to an intermediate phase existing in the field range from \SI{2.2}{\tesla} to \SI{2.4}{\tesla}. This intermediate phase has an inverted polarization that seems to be field-independent, although the limited field range prevents an unambiguous conclusion. If the polarization does not follow the linear magnetoelectric effect, this phase could be a distorted cycloid as in \pureNH\ and thus be a pseudo-multiferroic phase. Interestingly, this phase is stable up to the ordering temperature, which differs from the intermediate phase in the materials with less potassium content.
The stability of this intermediate phase indicates that the potassium concentration of 0.15 is close to the border between a multiferroic and a magnetoelectric ground state.
For $x=$ 0.45, see panel d), $T_\text{N}$ increases to \SI{10.8}{\kelvin}. As a function of increasing field there is only one spin-flop transition at $H_\text{SF}$. For both, $x=0.15$ and $0.45$, it is possible to switch the magnetoelectric domains for $|H| < |H_\text{SF}|$ with a magnetic field alone. Moreover, for all $x>0$, opposite single-domain states are realized in the respective spin-flop phases above positive and negative magnetic fields $H_\text{SF}$.
	
Figure \ref{diltion_overview} compares the substitution dependence of the N\'{e}el temperature ($T_\text{N}$), the Curie-Weiss temperature ($\theta$), and the structural transition temperature ($T_\text{S}$) of \NHK. By changing the potassium concentration $x$ we can tune the magnetic ordering temperature $T_\text{N}$ and the Curie-Weiss temperature, which both increase linearly with $x$, while the structural phase transition temperature $T_\text{S}$ decreases linearly.

\section{Conclusion}
Pyrocurrent measurements across the structural phase transition occurring in \NHK\ well above the magnetic order reveal the appearance of electric
polarization for finite K content while no such signal is observed in the pure material.
Using a deuterated crystal, neutron diffraction experiments characterize this phase transition as an order-disorder transition of the ND$^+_4$ groups.
In the high-temperature phase the ND$_4^+$ tetrahedra are disordered, which we model by the simultaneous occupation of a ND$_4^+$ tetrahedron orientation
and its inversion with respect to the central N position. Below the structural phase transition only one of these orientations is occupied at
each N site. This ordering breaks a glide plane resulting in the occurrence of special extinction rule violating reflections and the lattice becomes monoclinically distorted while the translation
symmetry does not change. Furthermore, a libration mode of the ordered NH$_4^+$ is observed by THz spectroscopy. The low-temperature structure is well described in a centrosymmetric space group, and further loss of this inversion symmetry or the
modeling in a non-centrosymmetric isotropy subgroup does not improve the description of the diffraction data. Considering the absence of electric
polarization appearing at this transition in pure \pureNH \ we can conclude that the low-temperature symmetry of the pure material remains non-polar.
The continuous emergence of polarization with increasing K content can be understood if the K insertion is not fully random but breaks inversion symmetry.
In this case the ordering of the NH$_4^+$ groups carrying a dipole moment can cause additional electric polarization; the materials with finite K content are thus pyroelectric.

The magnetic and dielectric properties of the substitution series \NHK\ systematically vary with $x$. The potassium concentration affects the magnetic exchange reducing the frustration (from $f=4.8$ for $x=0$ to 2.8 for $x=1$). Furthermore, the magnetic order transforms from a cycloidal ground state ($x\le0.06$) towards a collinear antiferromagnet ($x\ge0.15$) and, simultaneously, multiferroicity and spontaneous polarization are suppressed. The materials at higher K content exhibit linear magnetoelectric behavior. Due to the preexisting pyroelectric polarization it is possible to drive the linear-magnetoelectric antiferromagnetic domains without applying an electric field. Here, an external magnetic field is sufficient to switch between otherwise energetically degenerate states.

\begin{acknowledgments}
We acknowledge support by the DFG (German Research Foundation) via Project No. 277146847-CRC 1238 (Subprojects A02, B01, B02, and B04). The neutron data in this work were taken on the single crystal diffractometer HEiDi operated jointly by RWTH Aachen University and the J\"{u}lich Centre for Neutron Science (JCNS) within the JARA collaboration.
\end{acknowledgments}

\section{Appendix}

\begin{table}
\begin{tabular}{c c c c c c c}\hline
\multicolumn{2}{c}{$ 100\,\mathrm{K}, Pnma $}\\\hline
  & $ U_{11} $ & $ U_{22} $ & $ U_{33} $ & $ U_{12} $ & $ U_{13} $ & $ U_{23} $ \\\hline
 Cl1 & $ 12.7(4) $ & $ 5.0(4) $ & $ 14.5(6) $ & $ -0.6(3) $ & $ -2.6(4) $ & $ -0.5(3) $\\
 Cl2 & $ 10.9(6) $ & $ 9.7(6) $ & $ 13.6(8) $ & $ 0 $ & $ 3.3(5) $ & $ 0 $\\
 Cl3 & $ 9.3(5) $ & $ 10.6(6) $ & $ 8.6(8) $ & $ 0 $ & $ -0.9(5) $ & $ 0 $\\
 Cl4 & $ 12.1(6) $ & $ 12.4(6) $ & $ 9.2(9) $ & $ 0 $ & $ 1.3(5) $ & $ 0 $\\
 O1 & $ 14.1(9) $ & $ 9.8(9) $ & $ 21(1) $ & $ 0 $ & $ -9.5(9) $ & $ 0 $\\
 D1 & $ 27.4(7) $ & $ 17.9(7) $ & $ 30(1) $ & $ -4.6(6) $ & $ -8.4(7) $ & $ -3.7(7) $\\
 N1 & $ 14.0(5) $ & $ 13.2(5) $ & $ 13.2(7) $ & $ 0.8(4) $ & $ -0.5(4) $ & $ 2.1(5) $\\
 D2 & $ 55(2) $ & $ 67(2) $ & $ 16(2) $ & $ -4(2) $ & $ 1(1) $ & $ -2(2) $\\
 D3 & $ 50(2) $ & $ 22(1) $ & $ 60(3) $ & $ -13(1) $ & $ -12(2) $ & $ -6(1) $\\
 D4 & $ 23(1) $ & $ 55(2) $ & $ 43(2) $ & $ 0(1) $ & $ 4(1) $ & $ -4(2) $\\
 D5 & $ 36(1) $ & $ 23(1) $ & $ 61(3) $ & $ 7(1) $ & $ -7(1) $ & $ 11(1) $\\
\hline
~\\\hline
\multicolumn{2}{c}{$ 10\,\mathrm{K}, P2_1/a $}\\\hline
  & $ U_{11} $ & $ U_{22} $ & $ U_{33} $ & $ U_{12} $ & $ U_{13} $ & $ U_{23} $ \\\hline
 Cl1\_1 & $ 5.1(4) $ & $ 1.9(4) $ & $ 7.0(5) $ & $ -0.8(4) $ & $ -0.7(3) $ & $ 0.2(4) $\\
 Cl1\_2 & $ 6.2(4) $ & $ 2.5(4) $ & $ 4.7(5) $ & $ 0.0(4) $ & $ -0.2(3) $ & $ 0.4(4) $\\
 Cl2 & $ 4.4(3) $ & $ 4.3(4) $ & $ 5.2(4) $ & $ -0.7(4) $ & $ 1.0(3) $ & $ -1.0(4) $\\
 Cl3 & $ 4.9(3) $ & $ 4.5(4) $ & $ 4.5(4) $ & $ -0.7(4) $ & $ 0.0(3) $ & $ 0.4(5) $\\
 Cl4 & $ 5.7(4) $ & $ 3.9(4) $ & $ 5.0(5) $ & $ 0.3(4) $ & $ 0.8(3) $ & $ -0.2(4) $\\
 O1 & $ 6.6(4) $ & $ 5.7(5) $ & $ 9.2(6) $ & $ -1.2(8) $ & $ -3.6(5) $ & $ 2(1) $\\
 D1\_1 & $ 24(1) $ & $ 11.9(9) $ & $ 24(1) $ & $ -5.4(7) $ & $ -6.7(8) $ & $ -4.1(8) $\\
 D1\_2 & $ 22(1) $ & $ 14(1) $ & $ 30(1) $ & $ 3.2(7) $ & $ -8.5(9) $ & $ 2.3(8) $\\
 N1\_1 & $ 8.2(4) $ & $ 5.9(5) $ & $ 7.1(6) $ & $ -1.4(4) $ & $ -0.3(4) $ & $ -1.1(5) $\\
 N1\_2 & $ 8.1(4) $ & $ 6.2(5) $ & $ 8.1(6) $ & $ -0.7(4) $ & $ -0.4(4) $ & $ -0.6(5) $\\
 D2\_2 & $ 27.8(9) $ & $ 28(1) $ & $ 12.0(9) $ & $ -0.7(9) $ & $ 1.4(7) $ & $ -0.5(8) $\\
 D3\_2 & $ 36(1) $ & $ 16.4(9) $ & $ 29(1) $ & $ 7.3(8) $ & $ -6.6(9) $ & $ 3.6(8) $\\
 D4\_2 & $ 13.0(7) $ & $ 34(1) $ & $ 32(1) $ & $ 1.9(8) $ & $ 4.4(7) $ & $ -2(1) $\\
 D5\_2 & $ 25(1) $ & $ 16.4(8) $ & $ 26(1) $ & $ -6.0(8) $ & $ -6.0(8) $ & $ -5.5(8) $\\
 D6\_1 & $ 34.5(9) $ & $ 33(1) $ & $ 13(1) $ & $ -1(1) $ & $ 1.2(7) $ & $ -0.5(9) $\\
 D7\_1 & $ 26(1) $ & $ 14.9(9) $ & $ 36(1) $ & $ -7.3(8) $ & $ -3.9(8) $ & $ -6.1(9) $\\
 D8\_1 & $ 12.5(7) $ & $ 26.2(9) $ & $ 25(1) $ & $ -0.9(8) $ & $ 3.3(6) $ & $ -2.7(9) $\\
 D9\_1 & $ 25(1) $ & $ 17.0(9) $ & $ 31(1) $ & $ 6.8(8) $ & $ -4.0(8) $ & $ 6.2(8) $\\
\hline
\end{tabular}
\caption{Anisotropic displacement parameters of the two structural refinements with the data
taken at 100\,K and 10\,K on \ND ; all values are given in $10^{-3}$\AA$^2$.\label{tab:struc_parameters}}
\end{table}


\begin{thebibliography}{47}
\bibitem [{foo()}]{footnote}%
	\BibitemOpen
	\href@noop {} {}\bibinfo {note} {Note, that the settings $ P112_1/a $, $
		P2_1/n11 $, and $ P12_1/m1 $ correspond to $P2_1/a$, $P2_1/n$, and $P2_1/m$,
		respectively.}\BibitemShut {Stop}%
\end{thebibliography}

%

\end{document}